\documentclass[letter]{aa}

\usepackage{graphicx}
\usepackage{txfonts}
\usepackage{lipsum}
\usepackage{subcaption}
\usepackage{lscape}             
\usepackage{placeins}       
\usepackage[colorlinks,citecolor=blue]{hyperref}
\hypersetup{
  colorlinks = true,
  linkcolor = blue,
  anchorcolor = blue,
  citecolor = blue,
  filecolor = blue,
  urlcolor = blue}

\begin{document}

   \title{An extended red wing of the 5797\,{\AA} diffuse interstellar band\\ in the Red Rectangle nebula}

   \author{A. Ebenbichler\inst{1,2}\fnmsep\thanks{Corresponding author: aebenbic@uwo.ca}
        \and A.N. Witt\inst{3}    
        \and H. Van Winckel\inst{4}
        \and N. Przybilla\inst{5}
        \and P. Moraga Baez\inst{1,2}
        \and J. Cami\inst{1,2}
        \and E. Peeters\inst{1,2}
        \and A.~Candian\inst{6}
        \and I. Argyriou\inst{4}
        \and L.B.F.M. Waters\inst{8,9}
        }

   \institute{Department of Physics \& Astronomy, The University of Western Ontario, London, ON N6A 3K7, Canada
   \and Institute for Earth and Space Exploration, The University of Western Ontario, London, ON N6A 3K7, Canada
   \and Ritter Astrophysical Research Center, University of Toledo, Toledo, OH 43606, USA
   \and Institute of Astronomy, KU Leuven, Celestijnenlaan 200D, 3001 Leuven, Belgium
   \and Universit\"at Innsbruck, Institut f\"ur Astro- und Teilchenphysik, Technikerstr. 25/8, 6020 Innsbruck, Austria
   \and Anton Pannekoek Institute for Astronomy, University of Amsterdam, Science Park 904, 1098 XH, Amsterdam, The Netherlands
   \and School of Chemistry, University of Nottingham, University Park, Nottingham, NG7 2RD, United Kingdom
   \and SRON Netherlands Institute for Space Research, Niels Bohrweg 4, 2333 CA Leiden, The Netherlands
   \and Institute for Mathematics, Astrophysics and Particle Physics, Radboud University
   MC 62 NL-6500 GL Nijmegen, The Netherlands
   }

   \date{Received 12 September 2026 / Accepted 22 October 2026}
 
  \abstract
   {Diffuse interstellar bands (DIBs) are interstellar molecular absorption features that are ubiquitous in the interstellar medium (ISM).
   So far, there was only one known sight line where DIBs show an extend tail towards the red (ETR), namely Herschel~36.
   Such sight lines are essential to illuminate the DIB problem, namely the identification of their carriers, because they show more DIB variation than standard ISM sight lines.
   }
   {To report a new detection of the 5797\,\AA~DIB in the Red Rectangle nebula with an ETR, which adds a new observational constraint for the DIB carrier.
   } 
   {We analysed archival GIRAFFE/FLAMES integral field spectroscopy to extract all spectral features between 5776 and 5810\,{\AA} at different spatial positions.
   We used time series spectra to identify stellar and circumstellar features, and data from different instruments to exclude instrumental effects.}
   {We report the firm detection of the 5797\,\AA~DIB that is localised in the Red Rectangle nebula, or in the ISM in the sight line towards it.
   Curiously, while the 5797\,\AA~DIB shows an ETR in the Red Rectangle, the 5780\,\AA~DIB does not. Further, we report an exceptionally high equivalent ratio of $EW(5797)/EW(5780) = 1.79\pm0.13$, higher than any other literature value.
   }
   {}

   \keywords{Astrochemistry --
             ISM: individual objects: Red Rectangle --
             ISM: lines and bands --
             Line: profiles --
             Stars: AGB and post-AGB
             }

   \maketitle
\nolinenumbers

\section{Introduction}
The diffuse interstellar bands (DIBs) continue to be a mystery in astrochemistry since being discovered more than a century ago \citep{Heger22}.
They are ubiquitous in the Milky Way \citep{Zasowski15} and even in other galaxies \citep{HeLe00, Ehrenfreundetal02, Cordineretal08b}.
Besides the cationic fullerene C$_{60}^+$ \citep{2015Natur.523..322C,2017ApJ...846..168S}, no other DIB carrier is identified at this point.
A lot of progress was made, but the DIB carriers remain elusive.
Profile variation in DIBs is generally very subtle in sight lines through the diffuse ISM.
The only sight line with large profile variation is towards \object{Herschel 36}, the O-star illuminating the Hourglass region of the \object{Lagoon Nebula} \citep{2013ApJ...773...41D}.
Several strong DIBs show an extended tail towards the red (ETR), which was interpreted as rotational contours of highly excited carriers.
Anomalous sight lines like Herschel~36 are essential to solving the DIB puzzle, as they show extreme DIB profile variations.

In this Letter, we present a second sight line where at least one DIB has an ETR: the \object{Red Rectangle} nebula (RR).
The RR is a unique object for studying the formation and evolution of circumstellar and interstellar molecules in a broad wavelength range.
At its centre is the binary star system \object{HD 44179} consisting of a post-asymptotic giant branch (post-AGB) star that is losing mass and a solar-type main sequence companion \citep{2009witt}.
The resulting accretion disc around the secondary star ejects a bipolar mass outflow of dust and molecules, and is the source of ionising photons.
The binary system cannot be seen directly, but is obscured by a circumbinary dusty disk, where the dust scatters the optical star light, so the spectrum of HD~44179 is still observable \citep{2013MNRAS.430.1230T}.
The disk is extending outward more than 10 arcseconds and can be approximated by a spherical structure, only punctured by the ouflow \citep{2002A&A...393..867M}.
The outflow is a bright source of aromatic infrared bands (AIBs), and exhibits class B profiles in the centre \citep{2002A&A...390.1089P}, which is typical for circumstellar environments.
But it completely transforms to class A at larger separations \citep{2026A&A...709L..12E}, which is typical for the ISM.
In the optical, the outflow shows strong extended red emission (ERE), a very broad emission feature, giving the nebula its red colour \citep{1980ApJ...239L.133S}.
Additionally, there are narrow emission features superimposed on the ERE, called the red rectangle bands (RRBs).
Their peak wavelengths are very close to the wavelengths of some DIBs, leading to the assumption that they could be caused by DIB carriers in emission \citep{1992MNRAS.255P..11S,1995Sci...269..674S}. 
However, the absence of most DIBs in the central source HD~44179 \citep{2004ApJ...615..947H} casts doubt on this hypothesis.
Here, we report the detection of a 5797\,\AA~DIB with an ETR in the spectrum of HD~44179, which significantly enforces the idea of common DIB and RRB carriers and is only the second source of a DIB showing an ETR.

\section{Observations}
\begin{figure}
    \centering
    \vspace{-5mm}
\includegraphics[width=0.92\linewidth, trim={3cm 0cm 3.5cm 0.5cm}, clip]{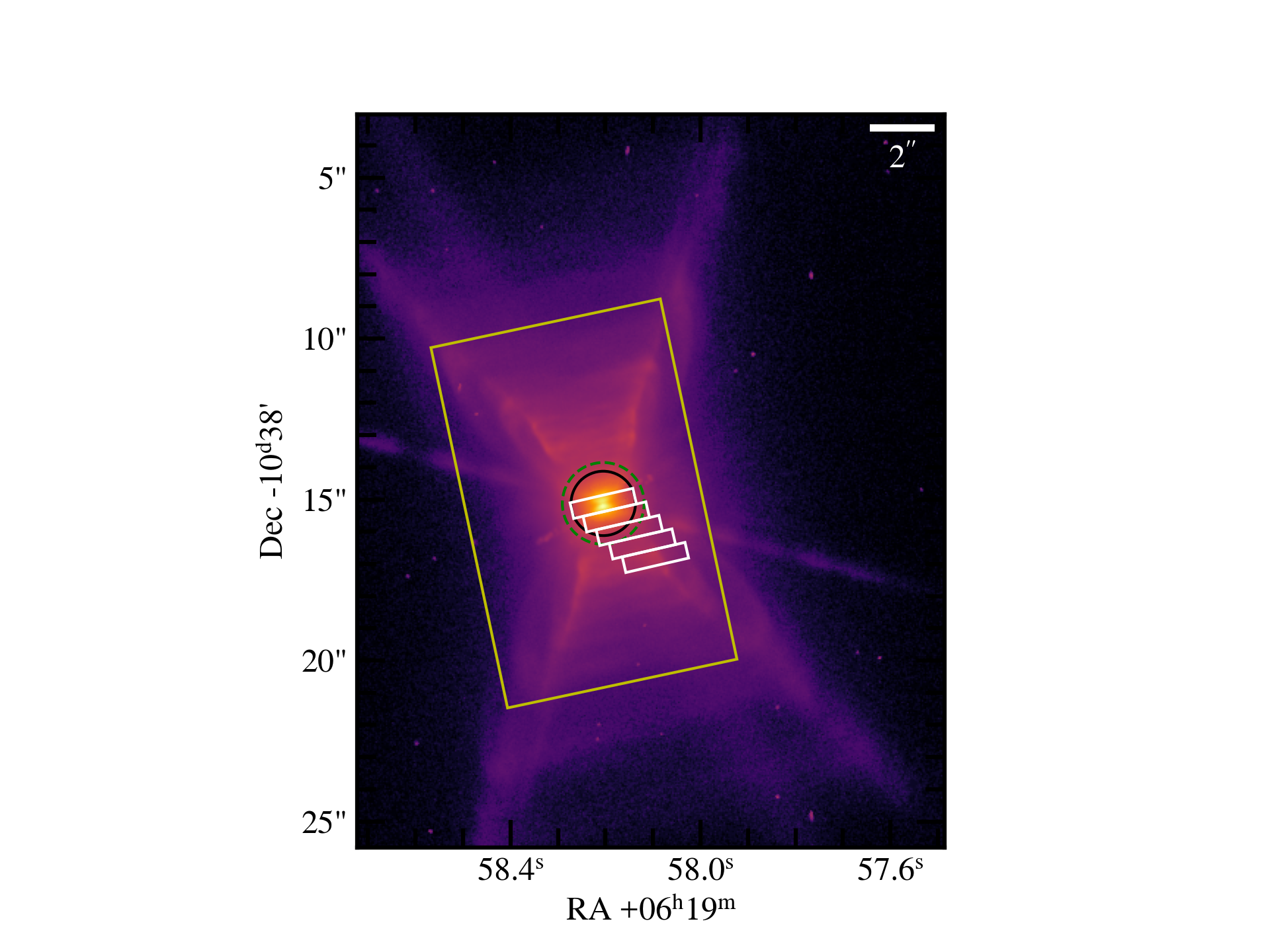}\\[-3mm]
    \caption{In yellow, the field of view of the FLAMES IFU data set on a broad band HST WFPC2 F622W image of the RR is indicated.
    The black circle shows the aperture used for extraction of data used in Figs.~\ref{fig:rr_herschel_hd185859} and \ref{fig:feros_check}, which coincides with the aperture of FEROS.
    The white apertures were used to extract data for the discussion in Fig.~\ref{fig:rr_centre}.
    The dashed green circle shows the aperture of the time series spectra, which are shown in Fig.~\ref{fig:time_series_spectra}.
    }
    \label{fig:flames_fov}
\end{figure}

We present integral field unit (IFU) data in the wavelength range of $5597 - 5840\,\AA$ and a spectral resolution of $\lambda / \Delta\lambda \equiv R = 37\,000$ taken with GIRAFFE in IFU mode on the Fibre Large Array Multi Element Spectrograph \citep[FLAMES,][]{FLAMESRef} on 30th of December 2004.
Further details of the observation can be found in \citet{2006ApJ...639..194S}.
From this point on, we refer to this as the FLAMES data.
We downloaded the raw FLAMES data from the ESO archive and reduced it using the ESO Data Processing System.
We re-projected the reduced cubes on a $0.2 \times 0.2$\arcsec grid, so a pixel-by-pixel comparison with JWST MIRI-MRS data is feasible, which is the basis of the main results in an upcoming publication (Ebenbichler et al., in prep).
To improve the astrometry, we calibrated the pointing of the FLAMES data to the central source, using a HST WFPC2 F622W image of the RR.
The resulting calibrated cubes were then coadded to get the final data product, which we shifted from the barycentric rest frame by a radial velocity of $-$21.5\,km\,s$^{-1}$ to the system rest frames of the central binary \citep{1995A&A...293L..25V}.
We extracted one dimensional spectra from the FLAMES data cube in multiple apertures (Fig.~\ref{fig:flames_fov}) to produce a collection of spectra at different separations from the central star HD~44179 (Fig.~\ref{fig:rr_centre}) and the central region (Fig.~\ref{fig:rr_herschel_hd185859}).

Additionally, we used spectral time series data from the HERMES spectrograph \citep{2011A&A...526A..69R} on the Mercator telescope in La Palma, Spain.
It is a fibre spectrograph with an extraction region of 2.54\,\arcsec~in diameter (Fig.~\ref{fig:flames_fov}) and a resolution of $R$\,=\,85\,000.
The time series consists of 235 individual spectra \citep{2018A&A...620A..85O} taken between October 20th 2009 and January 12th 2026 and phase folded (Fig.~\ref{fig:time_series_spectra}) with a spectral coverage of 3773 to 9006\,{\AA} and a mean $S/N$ of 57.
Just like the FLAMES data, the spectra were shifted to the system velocity rest frame of the HD~44179 binary system.
We chose the phase to be zero at JD 244\,8300 according to the definition of \citet{1996A&A...314L..17W} for easier comparison.

As complementary data, we downloaded two spectra of HD~44179 from the ESO science archive, observed consecutively on the 2nd and 3rd of May 2004 with the  Fiber-fed Extended Range Optical Spectrograph \citep[FEROS,][]{Kaufer99} on the MPG/ESO 2.2m telescope on La Silla in Chile. The fibrehead aperture covers 2.0\,\arcsec~in diameter on the sky. The spectra cover a useful wavelength range from about 4000 to 9000\,{\AA} at $R$\,$\approx$\,48\,000. After co-addition, a $S/N$ of 200 was estimated in the continuum near 5808\,{\AA}. 
For spectral data of Herschel~36, we co-added two spectra from the ESO science archive, observed on 15th of June 2020 with the Ultraviolet and Visual Echelle Spectrograph \citep[UVES,][]{UVES1} on the ESO VLT on Cerro Paranal in Chile. The spectra provide a wavelength coverage 5675 to 6646\,{\AA} at $R$\,$\approx$\,100\,000. After co-addition, a $S/N$ of 200 was measured in the continuum near 5808\,{\AA}. We applied a Doppler shift of 5.5\,km\,s$^{-1}$ to transform the spectra into the cloud rest-frame, using radial velocity measurements of the \ion{K}{i} absorption line.
In order to facilitate comparison with a standard DIB sight-line, we acquired UVES data of the early B-type supergiant \mbox{\object{HD 185859}} from the ESO science archive, observed on the 20th of September 2015 with coverage 4583 to 6686\,{\AA} at $R$\,=\,107\,200, and $S/N$\,$\approx$\,150. A Doppler shift of 7.5\,km\,s$^{-1}$ was applied, again using \ion{K}{i} as reference for the cloud rest frame. This sight line has one dominant cloud component and no peculiar broadened DIB profiles. In conclusion, all spectra used here were shifted to the cloud rest frame.

\section{Results\label{sec:results}}
\begin{figure}
    \centering
\includegraphics[width=0.9\linewidth]{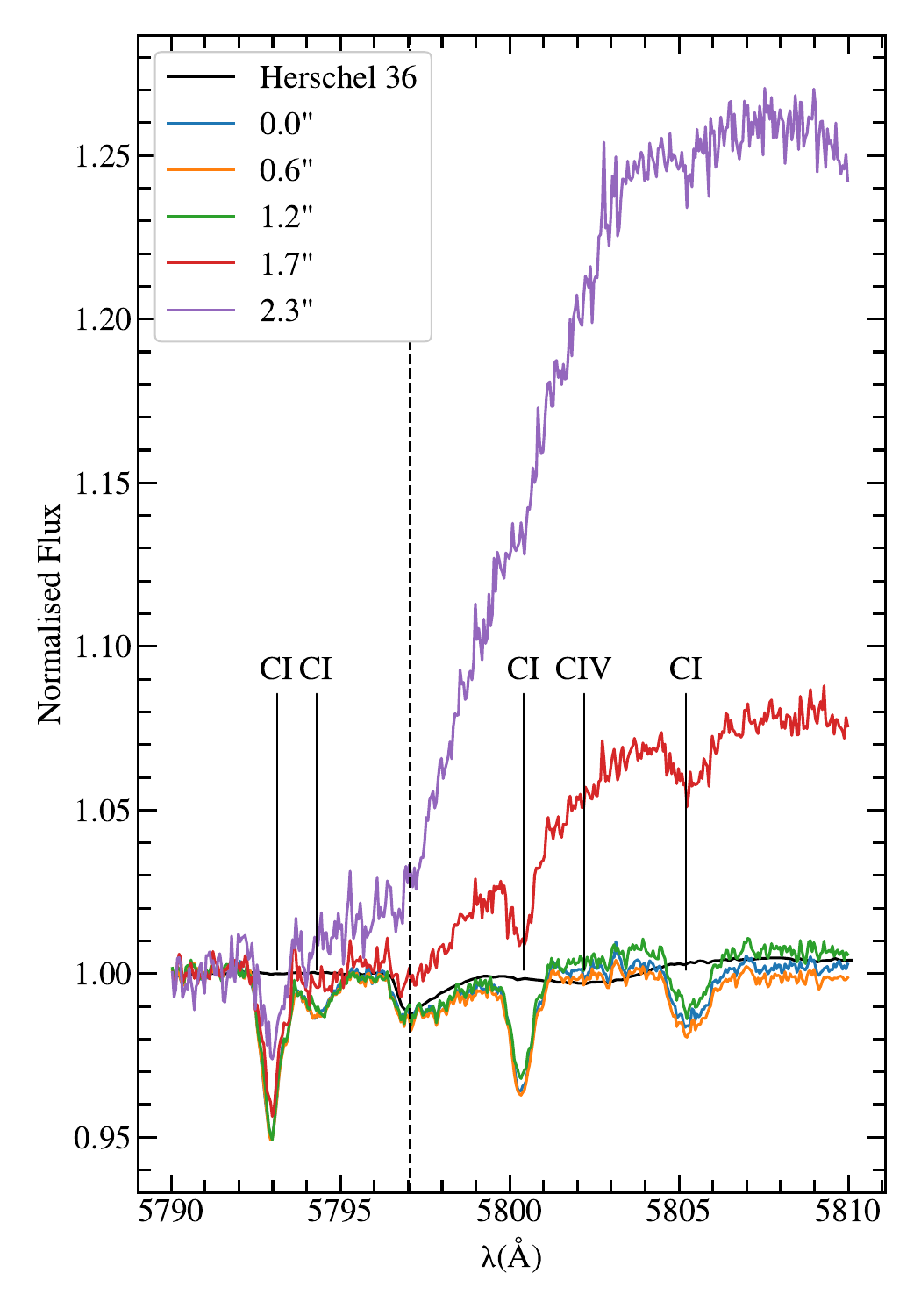}\\[-4mm]
    \caption{The 5797\,\AA~DIB in the central regions of the RR at different separations, normalised to the stellar continuum, compared to the spectrum in Herschel~36 which was stretched and degraded to match the DIB depth and resolution of the RR spectrum. Both spectra are in the respective cloud rest frame.
    Stellar carbon lines are annotated, while DIB wavelengths are indicated by vertical dashed lines.
    The \ion{C}{iv} line belongs to the O-type background star in Herschel~36.
    At larger separations, the RRB emerges as a broad emission right of the dashed line.
    }
    \label{fig:rr_centre}
\end{figure}

\begin{figure*}
\sidecaption   
\includegraphics[width=12cm, trim={0cm 0.5cm 0cm 0.5cm}, clip]{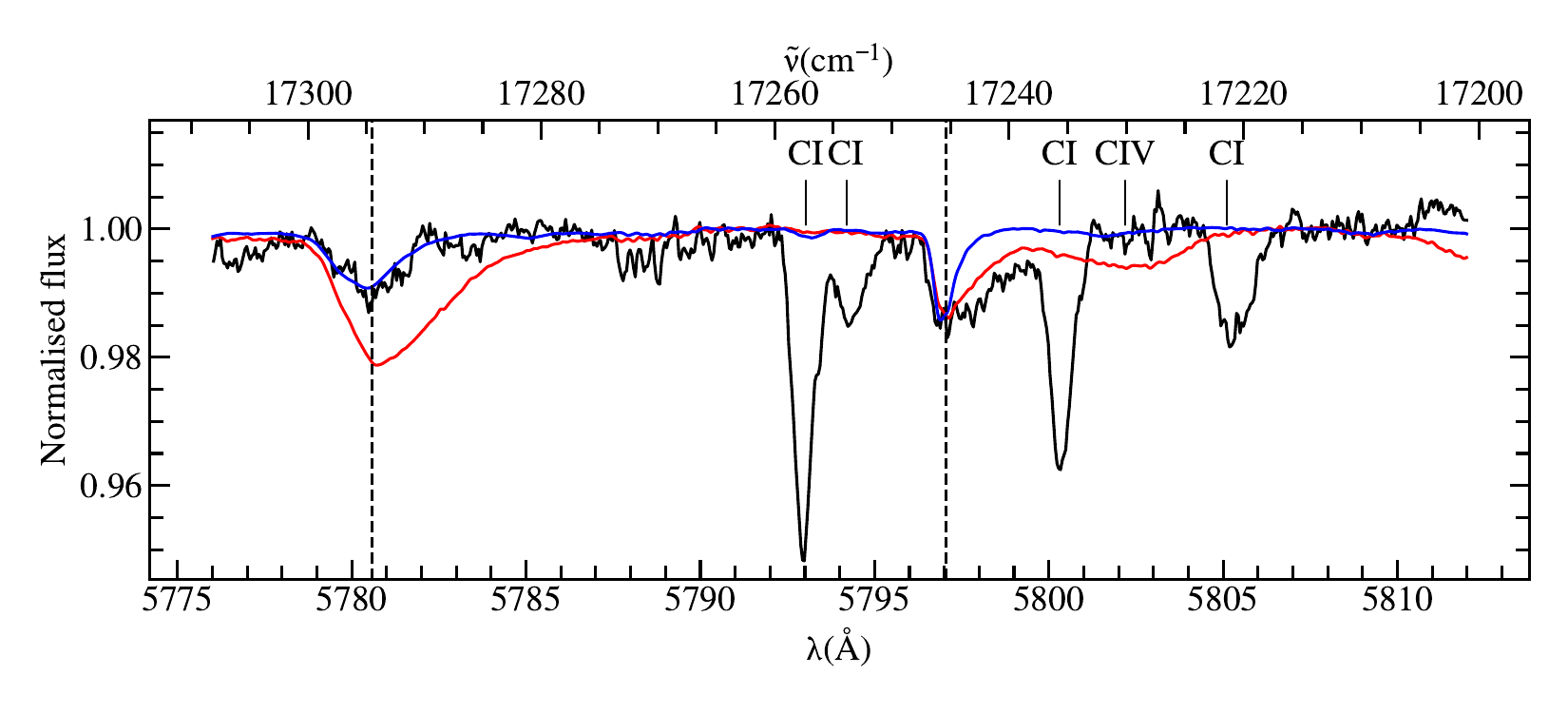}
    \caption{Comparison of the FLAMES DIB spectra in the RR (black), Herschel~36 (red) and the sight line towards HD~185859 (blue).
    We degraded the resolution of all spectra down to the lowest resolution of $R=37\,000$.
    The blue and red curves have been scaled, such that the 5797\,{\AA}~DIB has the same central depth.
    Stellar carbon lines are annotated, while DIB wavelengths are indicated by vertical dashed lines.
    The \ion{C}{iv} line is part of the stellar spectrum of the background O-type star in Herschel~36, and is also weakly seen in the early B-type supergiant HD~185859.
    }
    \label{fig:rr_herschel_hd185859}
\end{figure*}

\begin{figure}
\centering
\includegraphics[width=\linewidth, trim={0cm 0.5cm 0cm 0.5cm}, clip]{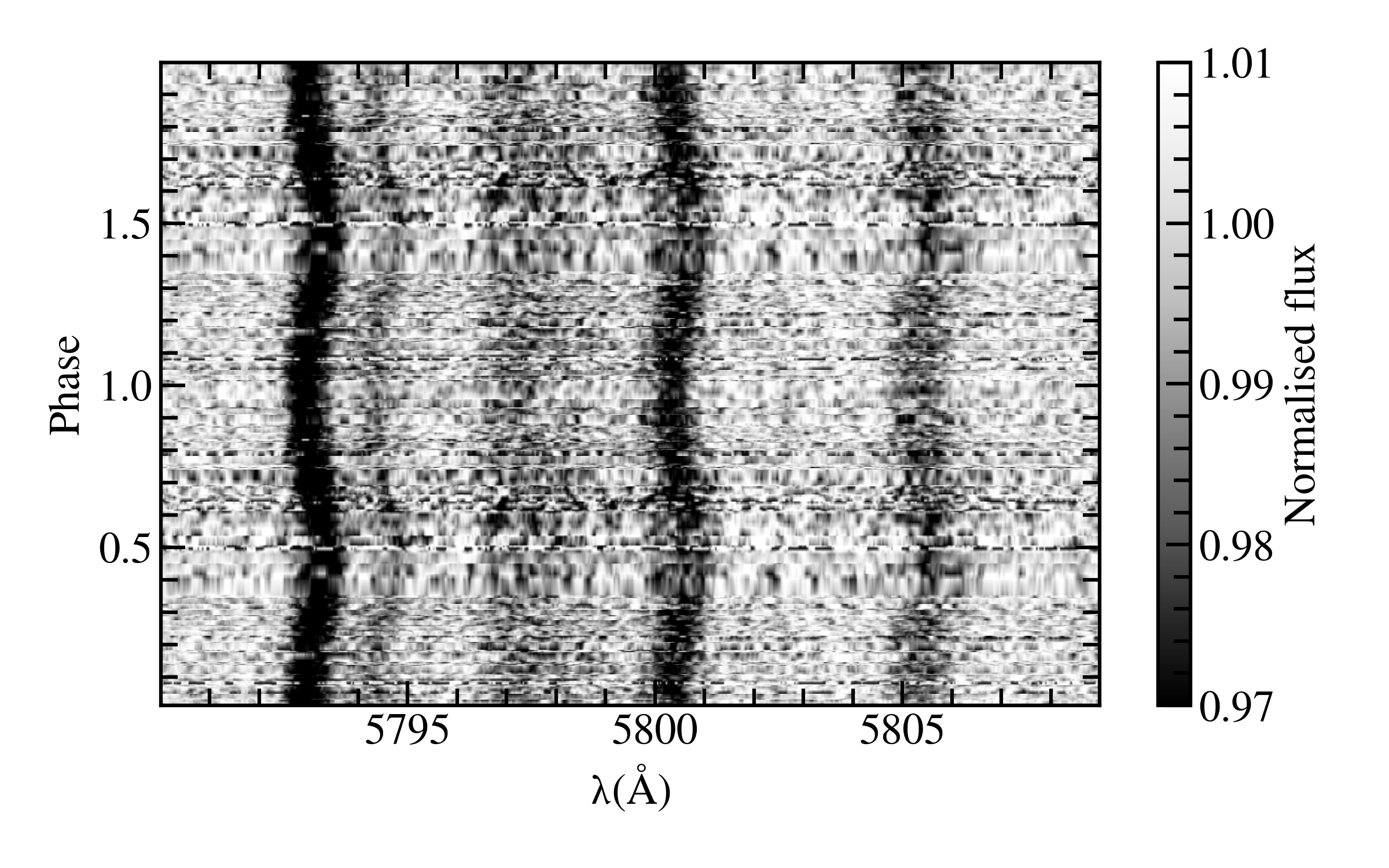}\\
    \caption{Optical time series spectra of the spectral region around the 5797\,\AA~DIB. 
    The rows were filled with 235 spectra which were smoothed with a box-car profile of 5 pixels.
    For the rows without data, the neighbouring spectra were interpolated.
    The stellar \ion{C}{i} lines are moving with the orbital phase of the primary star, but the broad DIB at 5797\,{\AA} is stationary, and does not change its band strength or profile shape. 
    }
    \label{fig:time_series_spectra}
\end{figure}

In the central region of our FLAMES data between 5790 and 5810\,{\AA} (Fig.~\ref{fig:rr_centre}, 0.0\,{\arcsec} and Fig.~\ref{fig:rr_herschel_hd185859}, black), we observe a stellar spectrum consisting only of \ion{C}{i} absorptions, with an additional DIB at 5797\,{\AA}.
The 5797\,{\AA}~DIB profile shape is very similar to the 5797\,{\AA}~DIB observed in Herschel~36 \citep{2013ApJ...773...41D}.
At a larger separation of 2.3\arcsec, the 5797\,{\AA}~DIB seems to get weaker, together with the scattered stellar lines (Fig.~\ref{fig:rr_centre}).
But because the S/N significantly decreases for larger separations, we cannot exclude the possibility of a 5797\,{\AA}~DIB there.
In Fig.~\ref{fig:time_series_spectra}, we see that the band strength and profile 5797\,{\AA}~DIB are  not changing significantly over the orbital phases.
At the same time, we see the 5800\,{\AA}~RRB emerge at 1.2\arcsec~separation, and getting much stronger relative to the scattered stellar continuum light at larger separations.
The blue wings of the RRB and the DIB start at the same wavelength of 5796.4\,{\AA}.
We measured equivalent widths of $EW(5780) = 20\pm1\,$m{\AA} and $EW(5797) = 36\pm2\,$m{\AA} using the error estimation of \citet{2011A&A...533A.129V}.
This results in an extremely high equivalent ratio of $EW(5797)/EW(5780) = 1.79\pm0.13$.
The full width at half maximum of the 5797\,\AA~DIB is FWHM$(5797)=2.5\,$\AA.

Before claiming a DIB detection, a potential confusion with stellar lines must be addressed.
From previous abundance analyses, we know that HD~44179 has a slight over abundance of carbon, standard abundances of nitrogen, oxygen, sulphur, and probably of zinc and the noble gases, and is generally under-abundant in refractory elements by order of 2-3\,dex \citep{Waelkensetal92,VanWinckeletal92}. 
In the spectral region between 5776 and 5810\,{\AA} in Fig.~\ref{fig:rr_herschel_hd185859}, only \ion{C}{i} lines are present in HD~44179.
The Doppler movement of the stellar \ion{C}{i} lines is well displayed in Fig.~\ref{fig:time_series_spectra}, while the DIB is not moving at all.
This behaviour definitely shows that the feature is a DIB and not a stellar line, so we can exclude the possibility of an unfortunate blend of stellar lines as the cause of this ETR absorption feature.
We can also exclude instrumental effects as the cause of the feature, as it is visible in both of our data sets, and also in additional FEROS spectra; see Appendix~\ref{sec:feros}.

The detection of a 5797\,{\AA}~DIB with an ETR is rather surprising, as the object has been studied multiple times in the context of DIBs \citep[see, e.g.][]{1995Sci...269..674S}.
However, optical studies mainly focused on the outflow of the nebula starting at a separation of 2.5\,\arcsec~\citep{2006ApJ...639..194S}, with less focus on the central region.
The central binary was studied by \citet{2004ApJ...615..947H} using the ARCES spectrograph with a resolution of $R$\,=\,38\,000 and a maximal $S/N$ of $850$.
They detected the 5780\,{\AA}~DIB and probably the 6613\,{\AA}~DIB, however nothing was detected at 5797\,{\AA}.
Further, at the time it was not known that some DIBs can show an ETR as it was discovered by \citet{2013ApJ...773...41D}.
Hence, a broad 5797\,\AA~DIB was maybe not recognised as such.

\section{Discussion}
The absence of Doppler movement in the time series spectra show that the 5797\,{\AA}~DIB is not photospheric, but of circum- or interstellar origin.
We cannot exclude an interstellar origin for certain, but there are several reasons why we think that a circumstellar origin is more likely.
Given the profile and ETR of the 5797\,{\AA}~DIB (FWHM = 2.5\,\AA), which is different than in other sight lines which probe the ISM, and its rest frame velocity, which coincides with that of the HD~44179 system, we favour the hypothesis that this 5797\,{\AA}~DIB absorption occurs in the immediate vicinity of HD~44179 in the RR itself rather than in the ISM along the line of sight toward Earth. 
Infrared pumping, as was discussed for the case of Herschel~36 \citep{Oka2013ApJ...773...42O} could very well be an important mechanism for the RR as well.
The circumbinary disk exhibits extremely high intensities of IR radiation \citep{2002A&A...393..867M}.
The intense infrared radiation field of the circumstellar environment may provide conditions favourable for radiative excitation of the 5797\,{\AA}~DIB carrier, potentially contributing to the observed extended red wing.
If the carrier was localised in the RR, this could give us the unique opportunity to determine the environmental conditions where the 5797\,{\AA}~DIB absorption arises and what leads to the enhanced red wing of the profile.

The detection of the 5797\,{\AA}~DIB in the HD~44179 spectrum, showing a pronounced ETR similar to that seen in Herschel~36, is very significant on several grounds. 
This is important in connection with the work of \citet{2019ApJ...886..138A}, who attempted to model the progression of profiles of the 5800\,{\AA}~RRB in the RR.
A crucial element of their model is the presence of a 5797\,{\AA}~DIB absorber within the RR, for which in 2019 there was no observational evidence. 
Our detection changes this, and it may now provide support for the \citet{2019ApJ...886..138A} model. 
Their model relies on the earlier paper by \citet{huang2015} that interprets the ETR in the 5797\,{\AA}~DIB profile in Herschel~36 as the rotational contour of a rotationally highly excited polar molecule with between 5 and 7 heavy atoms. 
A remaining weakness is that any molecule in this size range is not very robust in a radiation environment with far-UV photons. 
In the RR, one might argue that a short life time may be compensated for by rapid new formation of such a molecule to maintain a high abundance, but this may not apply to the low-density ISM, where 5797\,{\AA}~DIB is typically one of the strongest DIBs and where chemical reactions are much slower.
It can be seen in Fig.~\ref{fig:rr_centre}, that the blue edge of the 5796\,{\AA}~DIB and the blue edge of the 5800\,{\AA}~RRB coincide very well.
In the framework of rotational contours, those coinciding blue edges are compatible with the hypothesis that the DIB and the RRB are caused by the same carrier.

The standard shape of the 5780\,{\AA}~DIB, just like in the interstellar sight line towards HD~185859 is very surprising (Fig.~\ref{fig:rr_herschel_hd185859}).
It is in strong contrast to the 5797\,{\AA}~DIB (Fig.~\ref{fig:rr_centre}).
In Herschel~36, it has a pronounced ETR while the 5797\,{\AA}~DIB in Herschel~36 is narrower than in the RR.
The simultaneous ETR of the DIBs in Herschel~36 indicates that the excitation of the 5780 and 5797\,{\AA}~DIBs is governed by similar mechanisms.
The observations in the RR put this assumption to the test, as they behave very differently.
The different band shapes could reflect different carriers, different excitation conditions or physical environments.
The ETR for the 5780\,\AA~DIB could be caused by UV radiation which is much stronger in Herschel 36 than the RR.
Especially, the $EW$ ratio of $EW(5797)/EW(5780) = 1.79\pm0.13$ makes the RR sight line the most extreme $\zeta$-type sight line known in literature.
This value more than doubles the ratio in $\zeta$-Oph itself of $EW(5797)/EW(5780) = 0.80\pm0.04$ \citep{2011A&A...533A.129V}.
The high ratio might be influenced by a weaker UV-radiation field or a preferred formation or survival rate of the 5797\,\AA~DIB carrier.
If the DIBs in the RR are circumstellar, the formation could be influenced by the non-standard mixture in the RR, as HD~44179 is an evolved star.
To illuminate this question further, we will have to determine whether the DIBs in the RR are circum- or interstellar.

\section{Conclusion}
With the detection of the 5797\,\AA~DIB in the spectrum of HD~44179, there is now a second sight line where at least one DIB with an ETR has been observed.
The RR and Herschel~36 are essential parts in the DIB puzzle.
If the hypothesis that DIB absorptions in the RR arise within a confined region proves to be true, it could give us the opportunity to quantify the physical conditions, which are necessary to form their elusive carriers.
With the help of anomalous DIB sight lines, we can make significant progress on DIB research in the years to come.

\begin{acknowledgements}
      Part of this work was supported by \emph{ESO}, project
      number Ts~17/2--1. Based on data obtained from the ESO Science Archive Facility with DOI(s):
      \url{https://doi.org/10.18727/archive/24} and \url{https://doi.org/10.18727/archive/50}. 
      Based on observations collected at the European Southern Observatory under ESO programme 073.C-0337(A), 074.C-0025(A), 091.C-0851(A) and 194.C-0833(D).
      E. Peeters and J. Cami acknowledge support from the Canadian Space Agency (CSA, 24JWGO3A14), and the Natural Sciences and Engineering Research Council of Canada.
\end{acknowledgements}

\bibliographystyle{aa}
\bibliography{bibliography}
\listofobjects

\begin{appendix}
 \onecolumn

\section{Verification of DIB features with FEROS data}\label{sec:feros}
\begin{figure*}[ht!]
\sidecaption
    {\includegraphics[width=12cm]{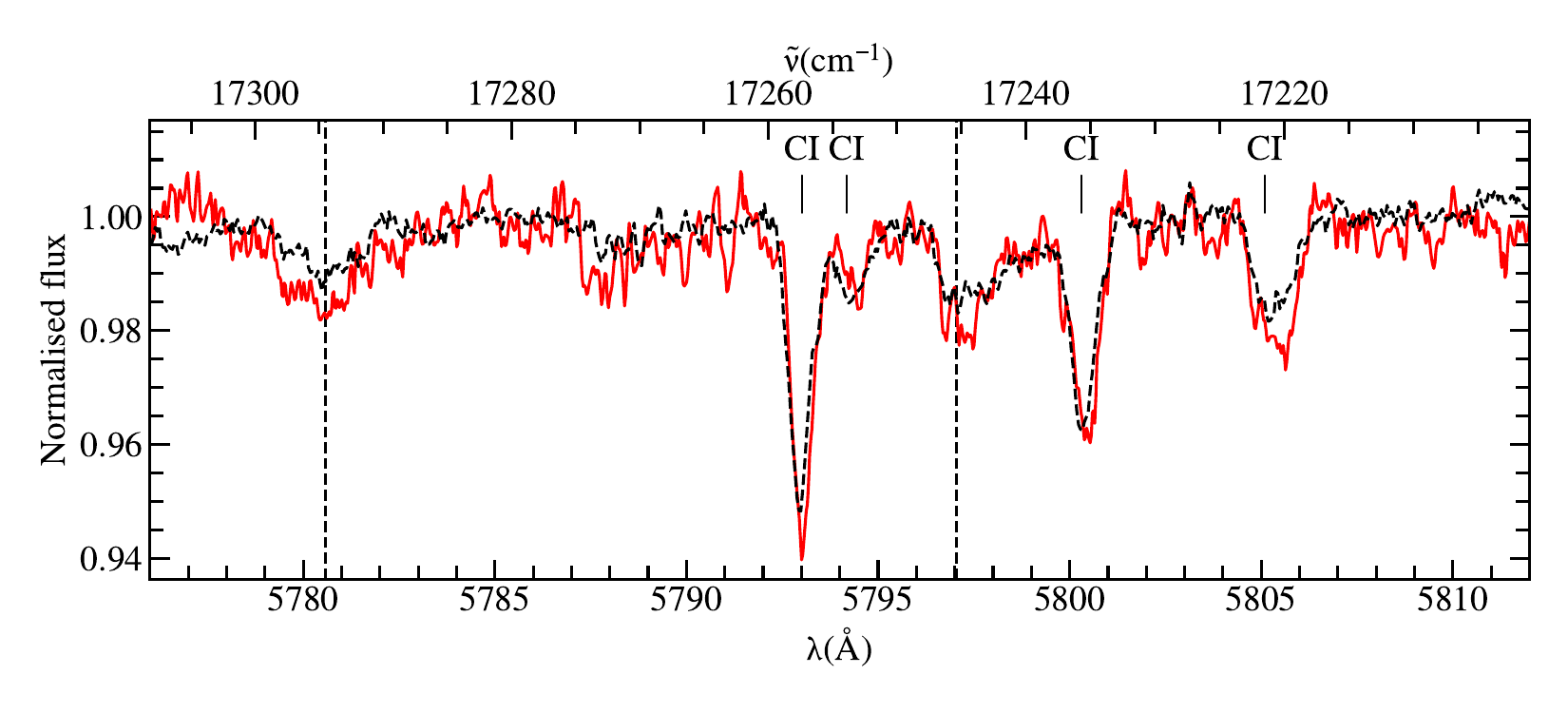}}
     \caption{Comparison of the FLAMES spectrum (dashed black) and the FEROS spectrum (solid red) of the central source.
     The two spectra agree within the noise level, and both show an absorption at 5797\,{\AA}.
     Stellar \ion{C}{i} lines are annotated, while DIB wavelengths are indicated by vertical dashed lines.
    }
      \label{fig:feros_check}
\end{figure*}
To make sure that the DIBs in the spectra recorded with FLAMES are not caused by instrumental effects, we compared the critical spectral region with the FEROS spectrum of HD~44179, see Fig.~\ref{fig:feros_check}.
Although the $S/N$ of the FEROS spectrum is lower and the resolution higher by about 30\%, we see that the 5797\,{\AA} DIB is similarly prevalent in both data sets. Hence, the features are not instrumental artifacts.

\end{appendix}
\end{document}